\documentclass[12pt]{article}
\usepackage{amsmath}

\newcommand{\be}{\begin{equation}}
\newcommand{\ee}{\end{equation}}
\usepackage{graphicx}

\def\eqnn#1#2{
	\begin{equation}#2\label{#1}
	\end{equation}}

\def\elnn#1#2{\begin{align}\begin{split}#2\label{#1}
	\end{split}\end{align}}
\def\elnnn#1{\begin{align}#1\end{align}}

\begin{document}
\baselineskip=24 pt

\begin{center}

{\large {\bf Perspectives on Galactic Dynamics via General Relativity}} 

\end{center}

\vskip1.5truecm 

\begin{center}
  F. I. Cooperstock and S. Tieu \\
{\small \it Department of Physics and Astronomy, University 
of Victoria} \\
{\small \it P.O. Box 3055, Victoria, B.C. V8W 3P6 (Canada)}\\

{\small \it e-mail addresses: cooperstock@phys.uvic.ca, stieu@uvic.ca}
\end{center} 

\begin{abstract} 
Responses to questions, comments and criticism of our recent 
paper ``General Relativity Resolves.." \cite{CT} are provided. It is 
emphasized that our model is entirely natural to describe the 
dynamics of an axially symmetric galaxy and that our solution, albeit 
idealized, contains the essence of the problem. The discontinuity of 
the metric derivative on the symmetry plane is necessarily 
interpreted as the effect of the mathematically idealized 
discontinuity of the gradient of the density and is shown to be 
naturally connected to the distributed volume density via the Gauss 
divergence theorem.  
We present arguments to the effect that for our approximate weak 
field model, we can choose the physically satisfactory mass
distribution without an accompanying singular mass surface
layer. To support this contention, 
we modify our solution slightly by removing the discontinuity with a 
region of continuous density gradient overlapping the $z=0$ plane. 
The alternative of invoking a surface layer leads to the presence of 
a negative mass surface layer approaching the numerical value of the 
positive mass continuous region. This is in contradiction with the 
assumed stationarity of the model. We find that a test particle behaves normally as it approaches the $z=0$ plane, the acceleration being towards the direction of this plane. This is in contradiction to the negative mass layer hypothesis as negative mass would repel the test particle. Thus, further support is added to the integrity of our original model. 
\end{abstract} 
\vspace*{1truecm}

\begin{center}

\textit{Subject headings}: galaxies: kinematics and 
dynamics-gravitation-relativity-dark matter\\
\mbox{ }\\
\end{center}

\section{Introduction}
Recently, \cite{CT} we presented a paper illustrating that 
Newtonian dynamics is inadequate to describe galactic dynamics. We 
showed that general relativity, the preferred theory of gravity, is 
required for the task even though the fields are weak and the motion 
is non-relativistic. This is because in such a gravitationally bound 
dynamical problem with an extended matter distribution, 
non-linearities cannot be neglected. We showed that general 
relativity allows the modeling of the essentially flat galactic 
rotation curves without exotic dark matter. In the process, we 
determined the mass density of the luminous threshold based upon data 
in the radial direction as given in \cite{kent}. In the short period since our presentation, 
we have received a large volume of correspondence with interesting 
questions, comments, suggestions and criticism (three such already 
posted \cite{korz}, \cite{VL} \cite{garf}) .
\footnote{
	It should be noted that in private communications, two 
	colleagues independently alerted us to the same line of
	reasoning as in \cite{korz} prior to that posting.
} 
The essential thrust of \cite{korz} was the claim that our particular 
model contained a singular disk of mass in the symmetry plane of the 
galaxy and thereby, vitiated our solution as a proper model for a galaxy. Subsequently, various individuals 
have used this argument to claim that our work is flawed. 
More recently \cite{VL}, it 
was suggested that the symmetry plane was the seat of exotic matter with one option being a negative mass sheet. Most recently, it was argued in \cite{garf} that the standard iterative perturbation scheme accounts for non-linearities and hence the galactic dynamics must be determined by Newtonian theory to lowest order. We will discuss these papers further in what follows.

In this paper, we respond to the criticisms. In the process, some 
new interesting insights emerge.  What must be emphasized is this: 
while it is certainly useful to have raised such criticisms, it is 
important to recognize that in the multitude of comments and 
correspondence that followed, \textit{no one to our knowledge has 
found valid reason to fault our central thesis, that general 
relativity, the preferred theory of gravity, exhibits essential 
non-linearities in sources of galactic scope}. While in subsequent 
studies, one solution to the model may be preferred over another, the 
essential point is that a new route to astrophysical dynamics is 
opened up by the recognition that general relativity, long accepted 
as the key to cosmology, also comes into play with significance for 
the major building blocks \textit{within} cosmology.

A frequent criticism of our work is the evidence that is presented 
for the existence of vast amounts of exotic dark matter in larger 
than galactic scales such as in the scale of clusters of galaxies. 
Two points must be made in this regard. Firstly, while some 
individuals have read into our work that we have claimed to have 
\textit{proved} that such large exotic dark matter reservoirs do not 
exist, this is not the case. Thus far our work applies to the 
galactic scale. Secondly, we have pointed to the fact that it would 
be interesting to extend the \textit{general relativistic} approach 
to the other relevant areas of astrophysics to determine whether or 
not exotic dark matter is truly required in those larger scale 
domains. For example, for the dynamics of clusters of galaxies, the 
virial theorem is used. This is based on Newtonian gravity theory. It 
would be of interest to introduce a general relativistic virial 
theorem for comparison. It is only after possible effects of general 
relativity are explored that we can be confident about the viability 
or non-viability of exotic dark matter in nature.

In Section 2, we review the essential structure and equations that 
were developed in \cite{CT}. In Section 3, we reply to certain issues 
that had been raised and in Section 4, we consider the problem of 
matter distribution. We end with concluding remarks in Section 5.

\section{Field Equations and Solution for Galactic Modeling}

We had modeled a galaxy in terms of its essential characteristics as 
a uniformly rotating fluid without pressure and symmetric about its 
axis of rotation. Within the context of general relativity, we began 
with the general metric structure
\eqnn{Eq1}{
	ds^2 =	-e^{\nu-w}( udz^2+dr^2)
		-r^2 e^{-w} d\phi^2+e^w(cdt-Nd\phi)^2
}
where $u$, $\nu$, $w$ and $N$ are functions of cylindrical polar 
coordinates $r$, $z$. To the order required, we found from the field
equations that $u$ could be taken to be unity.  As in previous
studies \cite{vs} \cite{Bonnor}, we chose to work in a coordinate
system co-moving with the matter having four-velocity 
\eqnn{Eq2}{
	U^i = {\delta}_0^i.
}
Actually, with this choice, it follows that $w=0$ as a result of the
requirement
\eqnn{Eq1ar}{
	g_{ik}U^iU^k =1.
}
As in \cite{Bonnor}, we performed a purely \textit{local} ($r$, $z$ 
held fixed) transformation
\footnote{
The importance of the nature of this transformation will be discussed in reference to (\ref{Eq50r}).
}
\eqnn{Eq3}{
	\bar{\phi} = \phi + \omega(r,z)\,t
}
that locally diagonalizes the metric. For the weak fields under 
consideration, this gave the approximate local angular velocity 
$\omega$ and the tangential velocity $V$ as
\elnn{Eq4}{
	\omega & \approx  \frac{Nc}{r^2},  \\
	V &=\omega r.
}
With $w=0$, the Einstein field equations reduce to
\elnn{Eq5}{
	2r\nu_r+ N_r^2-N_z^2 &=0, \\
	r\nu_z +N_r N_z &=0,  \\
	N_r^2 + N_z^2 +2r^2(\nu_{rr}+\nu_{zz}) &=0, \\
	\frac{3}{4}r^{-2} (N_r^2 + N_z^2)
 	+ Nr^{-2}\left(N_{rr} +N_{zz} -\frac{N_r}{r}\right)
	- \frac{1}{2}(\nu_{rr}+\nu_{zz})
	&= \frac{8{\pi}G\rho}{c^2}\\
}
to order $G^1$ and are readily combined to yield
\elnnn{
	N_{rr} + N_{zz} - \frac{N_r}{r} &=0 
	\label{Eq9a} \\
	Nr^{-2}\left(N_{rr} +N_{zz} -\frac{N_r}{r}\right)
	+ \frac{N_r^2 + N_z^2}{r^2} &= \frac{8{\pi}G\rho}{c^2}
	\label{Eq9b}
}
for $N$ and $\rho$. When (\ref{Eq9a}) applies
\footnote{
	While we would ordinarily set the first group of terms
	to $0$ in (\ref{Eq9b}) as a consequence of (\ref{Eq9a}),
	we retain it here
	for the purposes of later discussion.
},
the density is given by
\eqnn{Eq9c}{
	\frac{N_r^2 + N_z^2}{r^2} = \frac{8{\pi}G\rho}{c^2}.
}
$\nu$ is readily found from the remaining field equations. 

We should note at this point, as some colleagues have indicated, that 
the full Einstein equations bring in an additional factor of an 
exponential in $\nu$ in (\ref{Eq9b}). However this had been dropped in \cite{CT}
as its retention would induce corrections of order $G^2$. As well, it 
is worth reiterating the fact that since we are not dealing with the 
exact Einstein equations but rather with equations and solutions only 
up to first order in $G$, it would be wrong to expect the degree of 
mathematical precision that is generally applied to exact solutions.

As in other works, we expressed (\ref{Eq9a}) as
\eqnn{Eq10}{
	\nabla^2\Phi =0
}
where
\eqnn{Eq10a}{
	\Phi \equiv \int\frac{N}{r}dr.
} 
It is to be emphasized that this $\Phi$ is \textit{not} the potential 
of Newtonian theory.
\footnote{
Since $\Phi$ contains a partial integral, we could replace the RHS of (\ref{Eq10}) by an arbitrary function of $z$.
}
We have referred to $\Phi$ as the ``generating 
potential'' in \cite{CT}.  With (\ref{Eq4}) and (\ref{Eq10a}), we have the 
expression for the tangential velocity of the distribution in terms 
of the derivative of a harmonic function
\elnn{Eq11}{
	V&=c\frac{N}{r} \\
	&=c\frac{\partial {\Phi}}{\partial{r}}.
}
By separating variables, we were able to express the solution as a 
superposition of lowest-order Bessel functions with exponential factors
in the transverse dimension $z$ as \footnote{
	We thank B.R. Steadman for bringing to our attention the 
	paper by de Araujo and Wang \cite{AW} who consider as a 
	solution, one term of the form in the series (\ref{Eq13}). 
}
\eqnn{Eq13}{
	\Phi = \sum_{n}C_ne^{-k_n |z|}J_0(k_nr)
}
With (\ref{Eq13}) and (\ref{Eq4}), the tangential velocity is
\eqnn{Eq14}{
	V= -c\sum_{n} k_n C_n e^{-k_n |z|}J_1(k_nr).
}
We chose appropriate parameters in (\ref{Eq14}) to model the
observed galactic rotation curves and with (\ref{Eq11}) and (\ref{Eq9b}),
we deduced the corresponding galactic mass density distributions.
These were intuitively satisfactory distributions, indicating mass 
primarily concentrated about a disk-like configuration with 
diminishing density with increasing radius, as observed for the 
luminous matter. Moreover, the integrated masses were less than the 
published values using Newtonian theory and greater than the 
MOND \cite{bek} values. Thus, the observed galactic dynamics was in 
accord with our calculations without the need for massive extended 
halos of exotic dark matter that had been predicted by previous studies 
using Newtonian gravity theory. We noted that with this form of 
solution, the absolute value of $z$ had to be used to provide a 
proper reflection of the distribution for negative $z$, the matter 
below the central plane. This results in a discontinuity in $N_z$
at $z=0$ 
and as a consequence, in the strictess sense,
the solution is restricted to $z$ values different from 0.  The important point is this: the 
essential necessarily physical consequence of the use of $|z|$ is that the $z$ 
component of the \textit{gradient} of $\rho$ is discontinuous at 
$z=0$. Moreover, since the limits of the density are the same as the 
symmetry plane $z=0$ is approached from above or below, we could 
usefully define the value of $\rho$ at $z=0$ to have that limiting 
value. Thus, a globally continuous and finite density distribution 
source is established.
\footnote{
	See, however, the issues surrounding the concept of a
	singular layer of mass at $z=0$ in what follows.
}
This is our choice to properly model the physics of galaxies as
opposed to abstruse mathematical structures.
We will discuss this further in what follows.

\section{Replies to Certain Issues Raised}

There was a natural tendency for researchers to question the need for 
general relativistic analysis from the very outset. After all, the 
galactic gravitational fields are weak and the velocities of the 
stars are non-relativistic. In a variety of situations, it is indeed 
correct that weakness of field and slowness of velocity suffice to 
make Newtonian gravity theory an excellent approximation to physical 
phenomena, observed or predicted. In such situations, for example in 
planetary motion, there are only minute, albeit very interesting 
small changes to the orbits as a result of applying general 
relativity. Yet even in this, an essentially new phenomenon, the 
emission of gravitational waves, is predicted to occur within general 
relativity yet it is entirely absent in Newtonian gravity. Thus, it 
is helpful to keep an open mind with regard to forming conclusions as 
to what general relativity may yield in studies of dynamical sources.

In fact, even within the study of gravity waves, there are aspects of 
non-linearity that might not be expected upon first glance. Indeed, 
as Eddington had noted many years ago in the context of perturbative 
calculations for weak gravity waves generated by masses whose motions 
are driven by gravity itself (``gravitationally bound" or 
``free-fall" in the nomenclature of general relativity), 
non-linearities cannot be ignored. This was due to the fact that for 
gravitationally bound systems, the velocities are of the order 
$\sqrt{Gm/R}$ and hence non-linear terms that are ordinarily rejected 
in perturbation theory could be as large as the linear terms that are 
the usually sole terms that are retained. The novel aspect that we 
found is that the Eddington insight is also applicable for 
gravitationally bound \textit{stationary} time-independent rotational 
systems (i.e. systems not producing gravity waves) that exist in 
nature. We see this in (\ref{Eq9b}): The source $8\pi G\rho/c^2$ is 
equated to quadratic terms in $N$ and from (\ref{Eq4}), we see that 
$N$ is of order $G^{1/2}$ as a consequence of the gravitational bound 
aspect. Thus the consistency of the procedure is established. This 
encapsulates the essential departure of our approach from that of 
Newtonian gravity. 

It is correct that general relativistic effects are minute for the 
weak field gravitationally bound example of the solar system. But 
there, the dominant field is that of the sun and the planets are for 
most purposes properly treated as test particles in the solar field, 
guided by this field but not contributing to the global field. By 
contrast, in the galactic problem, the elements of matter are both 
guided by and essential contributors to the global field. By no means 
does this change the fact that the field is very weak but it does 
change the nature of the dynamics and the connection of field to 
source. Some have seen this effect as a peculiarity of our choice of 
co-moving coordinates. However, this is not the case: Other 
coordinates would have brought in an additional $\nabla^2 w$ into the 
equation for $\rho$, and $w$ which has a correspondence to the 
Newtonian potential, is of order $G^1$ as required. It is to be 
emphasized that the second non-linear term in $N$ in the $\rho$ 
equation (\ref{Eq9b}) cannot be removed globally by any choice of 
coordinates for our \textit{stationary} system and hence the 
non-linear aspect is essential. Linearized theory is simply 
inadequate to the task. 

\section{The Issue of Matter Distribution}

An issue first raised privately to us by some colleagues and later in 
 \cite{korz} \cite{VL}  concerns the nature of the matter 
distribution. They have noted that given the existence of the 
discontinuity of $N_z$ that we had pointed to in \cite{CT}, a 
significant surface tensor $S_i^k$ can be constructed with a surface 
density component given by
\eqnn{Eq1r}{
	(8\pi G/c^2) S_t^t = \frac{N[N_z]}{2r^2} -\frac{[\nu_z]}{2}
}
to order $G^1$. The notation $[..]$ denotes the jump over a 
discontinuity of the given function, here at $z=0$. Using 
(\ref{Eq5}), this becomes
\eqnn{Eq2r}{
	(8\pi G/c^2) S_t^t = \frac{N[N_z]}{2r^2} +\frac{N_r[N_z]}{2r}
}
It was claimed that this necessarily implied the existence of a singular 
\textit{physical} surface of mass in the galactic plane above and 
beyond the continuous mass distribution that we had found, thus 
rendering our model unphysical.

\begin{figure}
\begin{center}
\begin{tabular}{c c}
\begin{tabular}{c}
\includegraphics[width=2in]{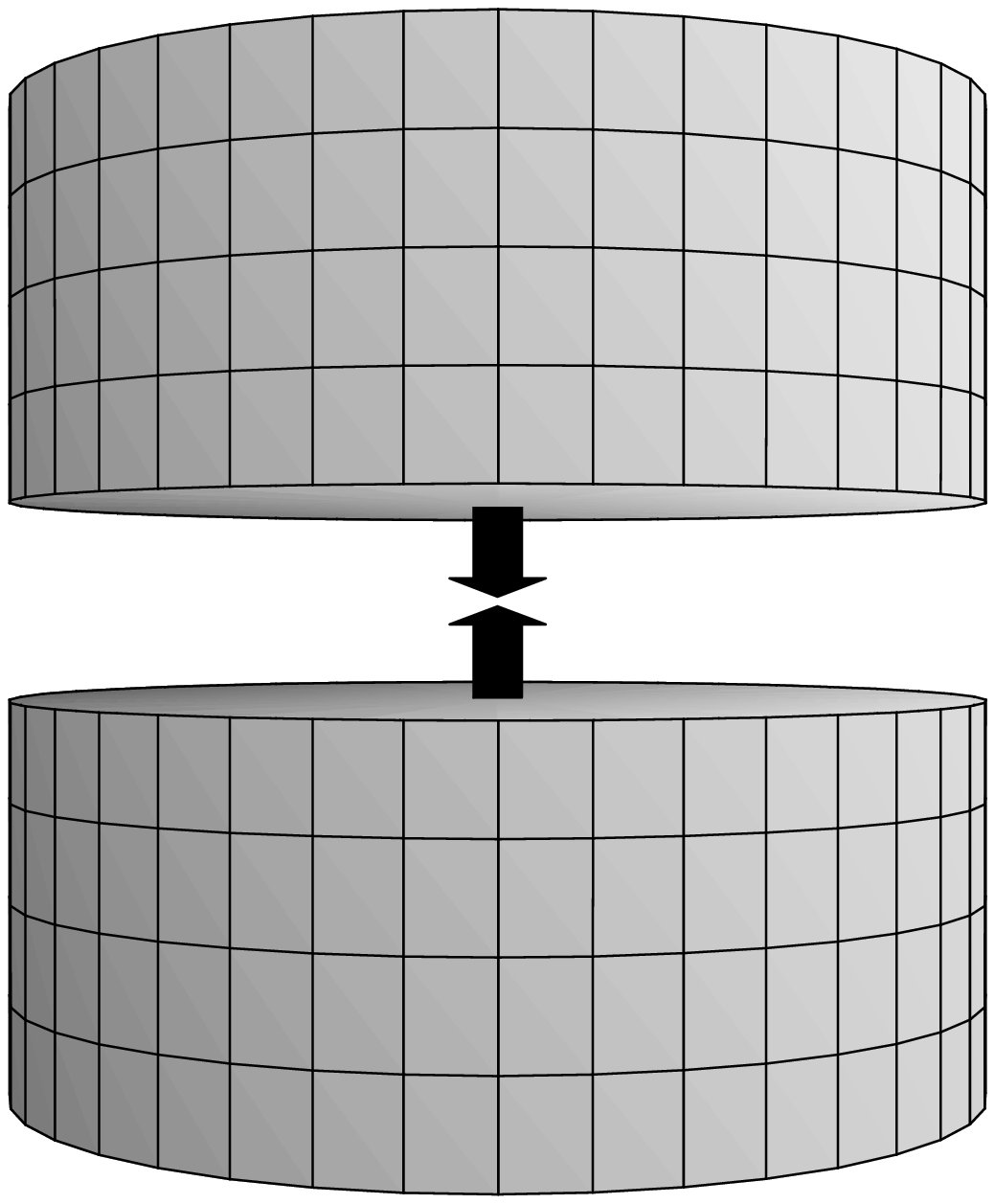}
\end{tabular}
&
\begin{tabular}{c}
\includegraphics[width=2in,height=0.75in]{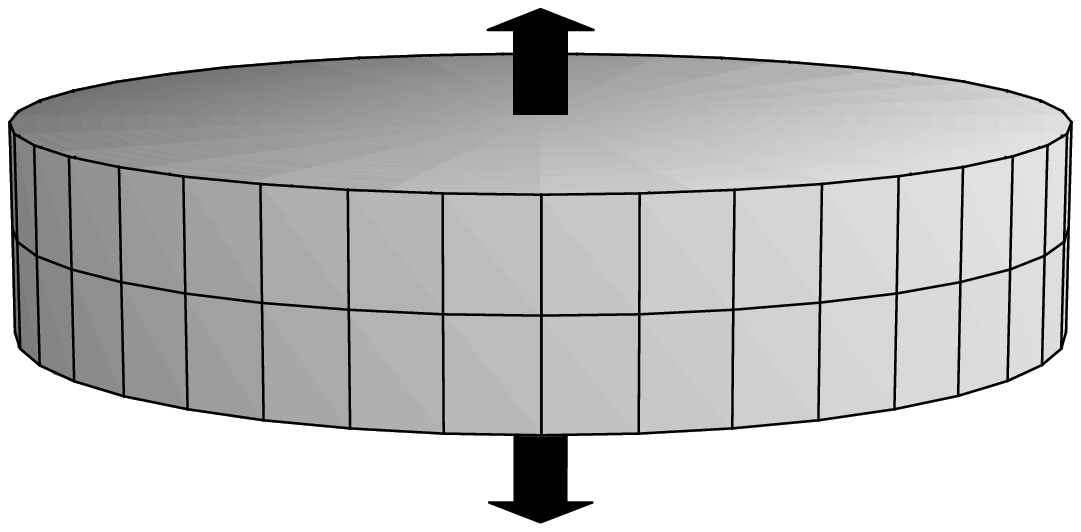}
\end{tabular}
\end{tabular}
\end{center}
\caption{
	Normal vectors used to calculate flux
}
\label{fig:cylinders}
\end{figure}

Having received this challenge, we calculated the surface mass that was said to be present in the four galaxies that we had studied by integrating (16) over the surface without paying 
heed to the actual sign of the result. Suspicions were aroused from 
the discovery that (\ref{Eq2r}) in each case gave a numerical value 
slightly less than the mass that we had derived from the volume 
integral of our \textit{continuous} mass density distribution using 
(\ref{Eq14}), (\ref{Eq11}) and (\ref{Eq9c}).
\footnote{
	It should be noted that the two terms in (\ref{Eq2r}) were
	found to contribute equally.
}
This pointed to a plausible explanation: in our case, \textit{with
our choice of model}, there is no \textit{physical} 
mass layer present on the $z=0$ plane. \textit{ The surface integral 
of this singular layer is merely a mathematical construct that 
indirectly describes most of the continuously distributed mass by 
means of the Gauss divergence theorem}. To see this, consider the 
vector ${\bf F}$ defined as
\footnote{
	${\bf e}_r$ and ${\bf e}_z$ are unit vectors in the $r$
	and $z$ directions.
}
\eqnn{Eq3r}{
	{\bf F} \equiv A(r,z){\bf e}_r + B(r,z){\bf e}_z
}
where 
\eqnn{Eq4r}{
	(8\pi G/c^2)B \equiv \frac{NN_z}{2r^2} +\frac{N_rN_z}{2r}
}
as a first option. We choose $A(r,z)$ so that 
\eqnn{Eq5ra}{
	\int\nabla\cdot {\bf F}dV \equiv (8\pi G/c^2)M
}
where $M$ is the total mass.
As a more transparent second option, we choose
\eqnn{Eq4ar}{
	(8\pi G/c^2)B \equiv \frac{NN_z}{r^2}
} 
where we define
\eqnn{Eq5r}{
	\nabla\cdot {\bf F} \equiv (8\pi G/c^2)\rho
}
From these definitions, we deduce
the form of $A(r,z)$ in order to produce the density as expressed 
through $N$ in (\ref{Eq9c}).  We calculate the mass over the 
cylindrical volume defined by $-\infty <z<\infty$, $0<r<r_{galaxy}$.  
By the Gauss divergence theorem, the volume integral of $\rho$, via 
(\ref{Eq5r}) is equal to the integral of the normal component of ${\bf F}$ 
over the bounding surfaces.  However, the integration must be over a 
continuous domain and since the ${\bf e}_z$ component is discontinuous 
over the $z=0$ plane, the volume integral must be split into an upper 
and a lower half. The two new surface integrals together would 
constitute the jump integral of (\ref{Eq2r}) in the first option if one were to be 
cavalier about the directions of unit \textit{outward} normals, as we 
shall discuss in what follows. The surfaces above and below the 
galaxy give zero because of the exponential factors in $z$ and the 
final small contribution comes from the cylinder wall via the $A$ 
function.

In our solution, the actual \textit{physical} distribution of mass is 
not in concentrated layers over bounding surfaces: the Gauss theorem 
gives the value of the \textit{distributed} mass via equivalent 
purely mathematical surface constructs as we are familiar from 
elementary applications of this theorem. Physically, the density is 
well defined and continuous throughout, except on the $z=0$ plane. In 
fact the limits as $z =0$ is approached give the same finite values 
from above and below. While the field equations break down at $z=0$, 
the density for a physically viable model is logically defined by
this limit at $z=0$. However, with the chosen form of solution, the 
density \textit{gradient} in the $z$ direction is discontinuous on the 
$z=0$ plane. This gradient undergoes a reversal for a galactic 
distribution with diminishing density in both directions away from 
the symmetry plane. It is most convenient to achieve this with an 
abrupt reversal as we have done.  There is no indication that this 
choice alters the essential physics.

\begin{figure}
\begin{center}
\includegraphics[height=3in]{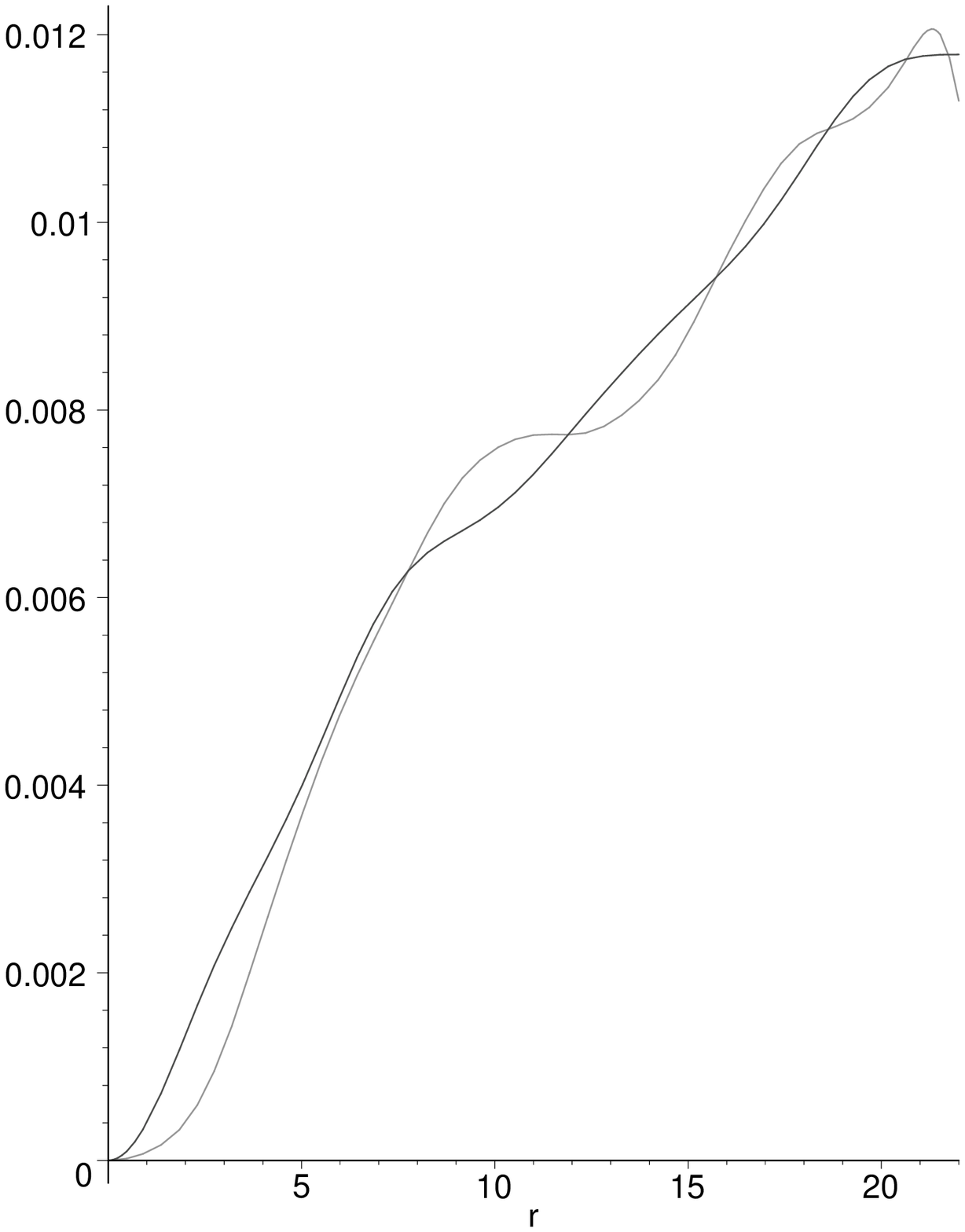}
\includegraphics[height=3in]{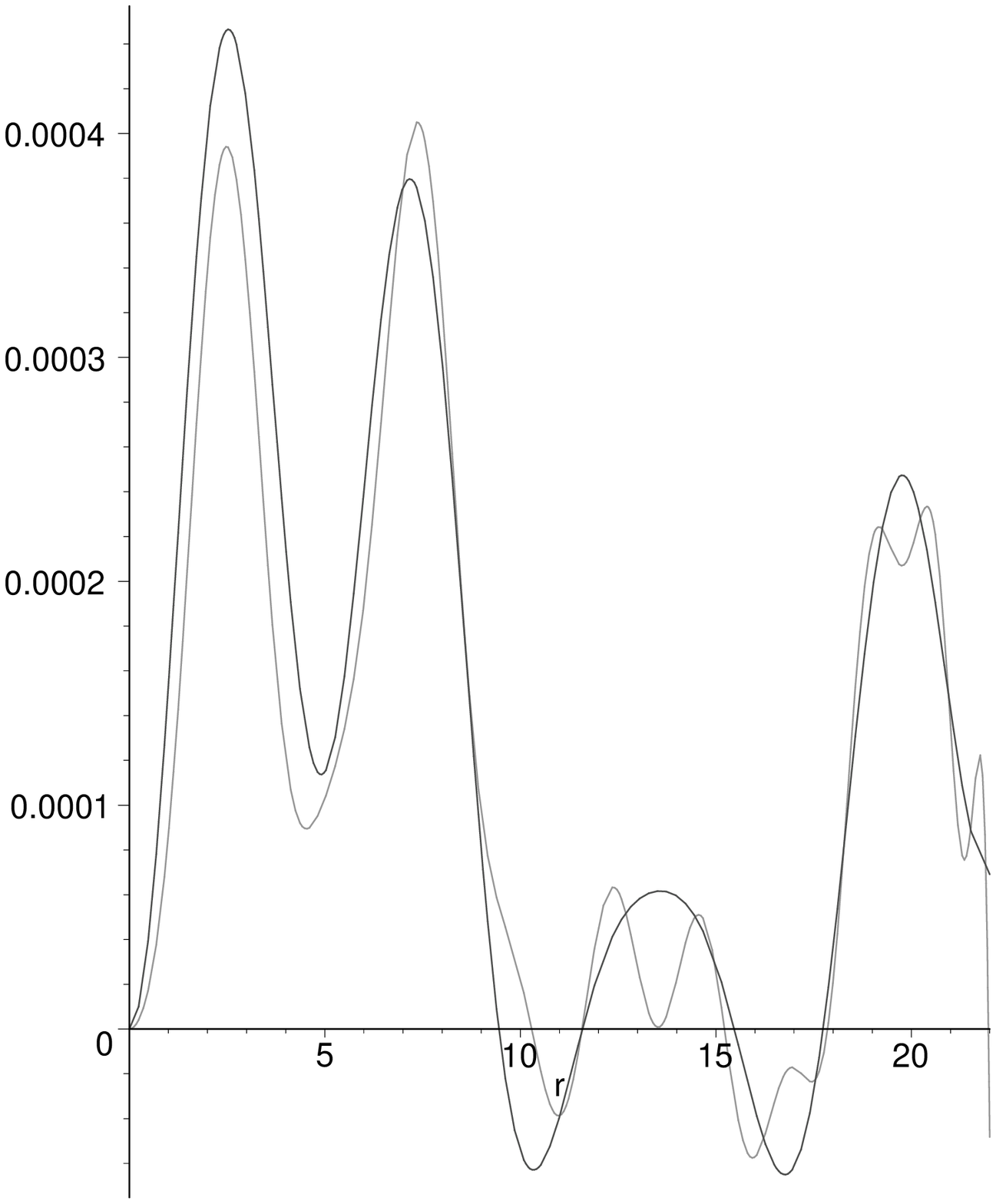}
\end{center}
\caption{
        Matching conditions for $N^{(ext)}=N^{(int)}$ and
        $N^{(ext)}_z= N^{(int)}_z$.
}
\label{fig:matching1000}
\end{figure}

To achieve the reversal with significant smoothness requires 
fine-tuning as we demonstrate in what follows.  Instead of 
exponential functions of the form that we chose in (\ref{Eq13}), we 
now choose combinations of exponentials \footnote{
	It is to be noted that we are not free to impose such 
	functions to suit our convenience, for example of the form 
	$e^{-kz^2}$ as suggested by some colleagues. While such 
	functions would exhibit exquisite smoothness and symmetry 
	about $z=0$, they would not satisfy (\ref{Eq10}) in the 
	separable form.
}
in the form of $\cosh(\kappa_n z)$ for $-z_0<z<z_0$ and in the original form 
of exponentials for $|z|>z_0$. By the choice of the $\cosh{\kappa_nz}$ 
functions to span the symmetry plane, we achieve smoothness over the 
interval that includes the $z=0$ plane and it is accomplished 
symmetrically about this plane. Clearly, since the values of
the $\cosh$ function and the exponential functions at $z=0$
\footnote{
	Strictly speaking, the matching to the rotational velocity 
	data could be seen to be effected at $z=\epsilon$ in order 
	to avoid the $z=0$ value.
}
are identical, the rotation curves now follow as before in \cite{CT}. 
However, this leads to the requirement that the functions $N$ and $N_z$ 
match at $|z| = z_0$. This was accomplished as is shown in Figure
\ref{fig:matching1000}. 
While this result is not easily achieved,
\footnote{
The careful addition of more terms in the series with proper choices of $\kappa_n$ and $C_n$ values would improve the fit further at the join.
}
it should be kept in 
mind that such difficulties likely stem from our restricting 
the solution class to sequences of \textit{separable} analytic 
functions in $z$ and $r$. Although these are most elegant to grasp and 
display, they invariably place restrictions on the forms of possible 
solutions. Had we resorted to direct numerical integration, the 
smoothing requirement issue about the $z=0$ plane would not have 
arisen. Moreover, had we opted for non-separable harmonic functions 
as solution generators such as sequences of Weyl line mass potentials 
\cite{BW}, this would also have averted the discontinuity issue. 
Interestingly, in its place thereby, we would have introduced a 
singularity at the origin. This would be pleasing to those who have 
suggested to us that it would be useful to model galaxies with black 
holes and/or naked singularities (see for example \cite{coop}) at the 
core, given the current suggested observational evidence.    

It is important to focus upon and be guided by the essential physical 
situation and not become ensnared in mathematical minutiae. The 
physical situation is one of a galaxy of stars freely gravitating in 
approximately  steady rotational motion. The model to describe it is 
a pressureless fluid with density symmetrically decreasing as we move 
vertically away from the $z=0$ plane in both directions. By way of 
contrast, consider the simplest example where (\ref{Eq2r}) comes into 
play in a directly \textit{physical} context: a spherical shell with 
Minkowski metric within and with Schwarzschild metric with parameter 
$M \neq 0$ outside. The field equations are for vacuum both inside 
and outside yet the spacetime has mass as evidenced by the 
Schwarzschild mass parameter. This indicates that there is indeed a 
\textit{necessarily} \textit{physical} layer of mass on the dividing 
shell as there is the necessity to have the mass manifest itself and 
there is no other place for it to reside. There is an essential 
physical discontinuity in this case in contrast to our rather benign 
galactic model density \textit{gradient} discontinuity: mass in the 
shell in the former but not necessarily on the plane in the latter.  
The argument in \cite{korz} treats our benign discontinuity (which we 
have seen is a density \textit{gradient} discontinuity) as a serious 
defect but we have seen that it has no necessarily essential effect 
on the physics. What is within our control is the decision as to 
whether we study a model with or without an actual singular layer of 
mass concentrated at $z=0$.  We have chosen the latter. There is no independent mass 
attributable to this discontinuity if we pursue this approach to 
develop a \textit{physically} viable model. It was also argued that 
our spacetime has irregular behavior asymptotically. However, in 
\cite{korz}, the harmonic coordinate conditions have been imposed in 
conjunction with the expansion that was used. 
By contrast, while we also deal with an expansion in powers of 
$G^{1/2}$, we had already exhausted all of the available gauge 
freedom in choosing the particular metric form with co-moving 
coordinate conditions; there is no room left for additional gauge 
restrictions such as the harmonic conditions.  Thus, the analysis in 
\cite{korz} does not apply to our situation. Moreover, we had noted 
in \cite{CT} that the Bessel functions fall off as $1/\sqrt{kr}$. 
Indeed, the authors in \cite{AW} have nicely presented the evidence 
in favour of this type of solution having asymptotic
flatness \footnote{
	This includes the fact that the Kretschmann scalar
	approaches zero for such solutions.
}
and noted some of the ambiguities associated with the issue.  

In a similar vein to \cite{korz}, \cite{garf} leads off with the well-known expression of the field equations in the harmonic
 gauge in Cartesian coordinates
\eqnn{Eq55r}{
     \partial_k\partial^kh^{ab}= \frac{16{\pi}G}{c^4}\tau^{ab}
}
where $\tau$ includes the energy-momentum tensor of the matter plus the non-linear terms in the Einstein equations. The standard description of the post-Newtonian perturbation scheme is invoked to conclude that the solution to the galactic problem must be the usual Newtonian one and that all corrections must be of higher order. However, what is unappreciated in the argument of \cite{garf} is that firstly we are not using this scheme, as we discussed above in conjunction with \cite{korz}, and secondly, that for gravitationally bound systems, the metric components are of different orders in $G$. 
\footnote{
Just as one would not logically choose Cartesian coordinates in the harmonic gauge to describe FRW cosmologies, one would not normally choose these for our stationary axially symmetric galactic problem. Our problem is greatly simplified with cylindrical polar coordinates co-moving with the matter. However, if one were to take the route as suggested in \cite{garf}, the equations could be schematically expressed as

$\nabla^2h_{(1/2)}=0$, $\nabla^2h_{(1)}= GT  + h_{(1/2)}^2$

\noindent
where tensorial superscripts have been suppressed and the lower case numbers refer to orders in $G$. In this manner, we would have incorporated the non-linear structure of our system within the framework of the scheme suggested by \cite{garf}. The novel aspect is that the lowest order equation (of order $G^{1/2}$) has zero on the RHS and the second equation that would normally be the Newtonian Poisson equation, differs in that it has non-linear terms.
}
Thus, the structure of our solution does not proceed as in (\ref{Eq55r}). In the latter, the lowest order base solution is the Newtonian solution whereas in the galactic problem, the lowest order equation for the density, (\ref{Eq9c}), has non-linear terms in the metric in the form of the squares of the derivatives of $N$. Thus, our situation is unlike standard iterative perturbation scheme applications as envisaged in \cite{garf}. Hence there is no basis to draw the conclusions that are expressed therein.

It is particularly illuminating to compare our model with the 
Newtonian Mestel disk model \cite{BT} \cite{mes}. In the latter, 
Mestel uses a solution of (\ref{Eq10}) in the form of functions used 
in (\ref{Eq13}). However, in this Mestel case, $\Phi$ is the 
\textit{Newtonian} potential and hence his solution in this context 
represents the physical condition of globally vanishing \textit{volume} 
density $\rho$ by virtue of the Poisson equation of Newtonian 
gravity. There is mass present as evidenced by the non-trivial 
potential that was selected. However, there is no place for it to 
reside except on the \textit{singular physical layer} at $z=0$. This is reminiscent of the situation with the Schwarzschild shell.

We contrast the Newtonian Mestel model with our general relativistic 
model. In both cases, solutions of the same type are chosen for 
(\ref{Eq10}). In the former, this implies vanishing volume mass density 
and hence an unambiguously physical surface mass layer. However, in 
the latter, a solution $\Phi$ of (\ref{Eq10}) does \textit{not} imply 
a vanishing volume density $\rho$. Our harmonic function plays a 
different role via general relativity from that of Mestel's harmonic 
function in Newtonian gravity theory. It is useful to think of 
general relativity as having the effect of opening up the Mestel 
disk, spreading out the mass continuously and symmetrically about the 
$z=0$ plane. By the choice of $|z|$ functions, the gradient of this 
dispersal is
necessarily discontinuous at the symmetry plane $z=0$. However, the 
mass is dispersed: the surface discontinuity that is present in this 
case is not necessarily
logically interpreted as an independent additional mass contribution. 
Indeed, let us examine this contribution more critically. Using the 
${\bf F}$ vector, we now evaluate the supposed mass that is harboured 
within the $z=0$ plane as seen in Figure \ref{fig:cylinders}. We do so by integrating the divergence of ${\bf F}$ 
over the volume from $z= -\epsilon$ to $z=+\epsilon$. As we see in 
the Appendix, the fact that the volume mass contributions that 
exclude the $z=0$ plane are positive both for $z>0$ and $z<0$ 
individually, the contribution from the surface layer is actually 
\textit{negative}
\footnote{
	In a private communication, W.B. Bonnor had informed us that 
	he had deduced that the mass layer would be negative. We 
	had come to this conclusion independently via the Gauss 
	theorem. In fact Bonnor had conjectured that the singularity at the origin     of his solution in \cite{Bonnor} had negative mass.
}
and integrations for the individual galaxies 
studied show the magnitudes to be close to those of the volume 
integrals \footnote{
	In fact, trial integrations for functions that are made to 
	vanish at the $r$ outer boundaries lead to perfect 
	agreement between the galactic volume integrals and the 
	absolute values of the surface integrals, as we would 
	expect from the Gauss theorem.
}.
In these results, we see two things: firstly, we see that the 
surface layer of mass is intimately connected to the volume 
integral of the clearly physical continuous mass density. Secondly, 
we see that if we were to pursue this to its logical conclusion with 
a refined distribution of density that tapers off to zero for very 
large $r$, we would be left with the patently false conclusion that 
the net galactic mass is zero. 

We would argue that the correct interpretation is this: in the case 
of a physically simulated continuous density distribution having a 
\textit{gradient} discontinuity as we have here, the $S_i^k$ tensor 
reflects this discontinuity and it does not necessarily signal 
additional mass or stress or angular momentum. 

Indeed this interpretation becomes more compelling when viewed as 
follows: suppose we were to begin in the more traditional physical 
manner, posit the energy-momentum distribution via the 
energy-momentum tensor and seek to solve the problem by solving for 
the corresponding metric. Let the $T^{00}$ component (i.e. $\rho$)
be given as
\footnote{
	For the purpose of illustration, we have used the numbers
	appropriate for the Milky Way \cite{CT}.
}
\elnn{Eq30r}{
	(8\pi G/c^2)\rho = &\frac{1}{r^2}\left\{\left(
	0.00126 r J_1(0.0687 r) e^{-0.0687 |z|} + \cdots\right)_{z}\right\}^2  \\
	&+\frac{1}{r^2}\left\{\left(0.00126 r J_1(0.0687 r) e^{-0.0687 |z|} +
	\cdots\right)_{r}\right\}^2 
}
After differentiating and squaring, this leads to a well-defined, finite, positive and continuous function for all points 
\textit{including} the $z=0$ plane. Indeed we take our density to have this value for all $(r,z)$, even for $z=0$. Since we are dealing with a pressureless 
fluid that is co-moving with the reference frame, all the remaining 
components of $T^{ik}$ are zero. As a result, the equation for 
$T^{02}$ yields (\ref{Eq9a}) everywhere, including the $z=0$ plane. 
From (\ref{Eq9b}), this yields
\eqnn{Eq31r}{
	\frac{N_r^2 + N_z^2}{r^2} = \frac{8{\pi}G\rho}{c^2}
}
as in (\ref{Eq9c}) where the $\rho$ is the posited one of 
(\ref{Eq30r}). \textit{Clearly the solution is as before only through 
this route, we do not have an additional surface layer of matter}. In 
effect, this is the approach that we followed in \cite{CT}. The 
difference is that in \cite{CT}, we approached the problem from the 
opposite direction, but the intent was the \textit{physical} model as 
developed here.

By contrast, we could have followed a variation of this approach as 
pursued, in effect, in \cite{korz} \cite{VL} \cite{AW}. We could have 
posited $\rho$ as in (\ref{Eq30r}) for $z \neq 0$ and having a 
singular layer at $z=0$. In this case, we would have taken $N_{zz}$ 
as a delta function for $z=0$ and the component $T^{02}$ would have 
had a delta function layer as well at $z=0$. In this case, 
(\ref{Eq9b}) would have been the operative equation for the density, 
now displaying the layer through the $N_{zz}$ term now present. 
However, this leads, as we have seen in the Appendix, to the 
production of a \textit{negative} density layer with almost the same 
numerical value as the positive mass continuous contribution.

As Bondi had noted in his writings, negative mass repels all masses, 
whether they be positive or negative. Thus, this enormous storage of 
negative mass on the plane could not maintain the envisaged 
stationary distribution and would blow itself apart. This underlines 
the problem associated with assuming that the $S_i^k$ tensor will 
necessarily indicate a viable expression for physical mass in all 
situations. In our case, it leads to an untenable model yet at its 
source, the structure actually arises in physical terms from a rather 
benign density \textit{gradient} discontinuity.     

In this regard, it is of interest to consider the behaviour of a test particle having the angular velocity and radial velocity of the bulk fluid but with a non-vanishing transverse velocity $dU^z/ds$ in the co-moving frame. The geodesic equation in the $z$ direction reduces to  
\eqnn{Eq50r}{
    \frac{dU^z}{ds}= \frac{N_rN_z(U^z)^2}{2r}
}
We computed the complete $N$ series for NGC7331 for $r=0.1$ to $30$ and $z=0.001$ to $1$ for the right hand side of (\ref{Eq50r}). All of the points gave a negative value as expected for the acceleration of a particle approaching a normal $z=0$ boundary from above. However, if the $z=0$ surface actually harboured a \textit{physical} negative mass surface layer, indeed one of numerical value comparable to the positive mass of the normal galactic distribution,
\footnote{
This is assuming that a mechanism could be found to prevent the layer from exploding.
}
then one would have expected to witness a violent repulsion of the particle as the test particle approached the boundary. The absence of this occurrence adds further support to the integrity of our original model \cite{CT}. 
\footnote{
In an interesting recent paper \cite{bon2}, the motion of a test particle for a different type of distribution was analyzed in the locally non-rotating frame produced via the transformation in (\ref{Eq3}). In this case, the $z$ geodesic equation is dominated by  $\Gamma^z_{00}$ for non-relativistic particles. In \cite{bon2}, this term indicated that $dU^z/ds$ was positive for $z>0$ and hence implied an apparent repulsion of the test particle. However, the geodesic equation should apply to particles of the dust itself since they are geodesic. These particles are at rest in the original frame and have only tangential velocity in the locally non-rotating frame. The local transformation should not alter their having no $z$ velocity yet there is an apparent $z$ acceleration. 

This is a contradiction which is resolved as follows: while we can use the local transformation to derive the local angular velocity (and hence tangential velocity) of the particles, it is not legitimate to take derivatives of the metric found after the local transformation has been applied in order to derive acceleration. The former usage simply reads off the required angular velocity to diagonalize the metric locally. No differentiation is required to do so. However, the latter usage would be legitimate only if the transformation would have been effected \textit{without} constraints, in this case the constraint of holding $r$ and $z$ fixed. By holding these fixed to derive the new $ds^2$, we have metric tensor components that are correct as such only through a different transformation at each point. Hence there is an inherent discontinuity of transformation. In this manner, we recognize the derivative required to find $dU^z/ds$ as being illegitimately applied, thus resolving the contradiction. 

This also brings into question the interpretation of the lack of a $1/r$ term in $g_{00}$ in the locally non-rotating frame for the same reason. The $1/r$ issue is a global one yet the transformation that brought the metric to the form that is being used is a purely local one. By contrast, no such problem arises in studying test particle motion relative to the dust co-moving frame with the particle following the motion of the dust cloud apart from a $z$ velocity component. 
The result is logical: for $U^r$ and $U^{\phi}$ being zero, acceleration occurs only for $U^z$ different from zero (otherwise it would be part of the dust cloud and hence stationary) and the acceleration is independent of the sign of the velocity. Moreover, the acceleration is \textit{negative} for $z>0$ and \textit{positive} for $z<0$. Thus, it is \textit{attracted} to the central plane in both cases.

We thank Professor W.B. Bonnor for bringing his paper to our attention.
} 

It is to be emphasized that the entire issue of singularities arose 
simply because of the wish to maintain the simplicity of solution 
with a sequence of exponential functions. The only irregularity that 
this necessarily entails is within the density \textit{gradient} and 
not necessarily within the density itself. We have shown that even 
this irregularity can be averted by the introduction of a $\cosh$ 
series. However, this leads to the challenge of matching interior and 
exterior regions mathematically. It should be stressed that while 
this is mathematically challenging, the actual physical distinction 
is minimal: the density gradient discontinuity, a mathematical 
idealization, is not present physically. In actual physical terms, the density is necessarily rounded out to some extent, however abruptly. It is never 
infinitely sharp.

\section{Summary and Concluding Remarks}

In \cite{CT}, we had modeled a galaxy as an axially symmetric pressureless 
stationary rotating fluid within the framework of general relativity 
to order $G$. We had shown that the dynamics was driven by one linear 
and one non-linear equation as opposed to the linear equation of 
Newtonian gravity. A framework of solutions with separated variables 
was established as a sequence of Bessel functions. This enabled us to 
choose appropriate parameters to fit the flat galactic rotation 
curves without invoking massive halos of exotic dark matter as is 
required using Newtonian gravity. The masses were concentrated 
primarily within the disk configuration. The mass values were found 
to be between those of Newtonian gravity and those predicted by the 
MOND model.

The separated variable approach led to exponential dependence in the 
transverse $z$ variable and reflection symmetry implied a 
discontinuity in density gradient at the symmetry plane. Critics had 
claimed that this necessitated an accompanying singular mass layer on 
this plane. 

In this paper, we approached the problem in two ways: firstly, we 
took the normal route of specifying the source as a singularity-free 
density distribution with density gradient discontinuity and saw that this led to the solution that we 
had used in \cite{CT}. The key was the observation that the field 
equations only dealt with $z$ derivatives in a form that led to 
unique limits as one approached the symmetry plane from above or 
below. This enabled us to specify the values of the functions on the 
plane as the limit as the plane was approached, from above or below.

Secondly, we bridged the region of the symmetry plane where the 
density gradient was discontinuous by using a new solution there, one 
that was symmetric and smooth, in fact infinitely differentiable. We 
showed that this solution could be metric and metric derivative 
matched to the original satisfactory exponential fall-off exterior 
solution. 

We also considered the consequences of \textit{not} demanding a 
natural non-singular density distribution as was invoked in 
\cite{korz} \cite{VL} \cite{AW}.  In this case, we found that the mass of the layer was necessarily connected to the mass of the continuum by the Gauss 
divergence theorem. In turn, this implied a negative mass in the 
layer that numerically approximated the positive mass of the 
continuum. Such an enormous negative mass would contradict the 
assumed stationarity of the model. Finally, we considered the motion of a test mass approaching the $z=0$ plane, thoroughly searching the domain for any sign of repulsion as would be implied by the supposed vast store of negative mass in the $z=0$ plane (assuming that the matter in the plane could somehow actually hold itself together). None was found in contradiction to the layer assumption.

In future work, it will be of interest to refine the process further 
with finer resolution of sequence solutions and with applications to 
more galaxies. It would also be of interest to explore the untapped 
area of non-separable solutions with their singular cores. These 
might be particularly useful should it be thoroughly established that 
galactic cores harbour physical singularities, be they clothed or 
naked.

\vskip 0.25in

{\small {\bf Acknowledgments:} We are grateful to W.B. Bonnor, F.D.A. 
Hartwick, D.N. Vollick and many colleagues for their valuable 
comments, suggestions, questions and criticism. This work was 
supported in part by a grant from the Natural Sciences and 
Engineering Research Council of Canada.}

\clearpage
{\small 

\section{Appendix}

From (\ref{Eq3r}) and the Gauss theorem,
\eqnn{Eq40r}{
	\int \nabla\cdot{\bf F} dV
	= \int {\bf F}\cdot{\bf n}dS
	=\int \rho dV.
}
Concentrating first on the upper cylinder, the surface integral 
contribution at $z=+\infty$ gives zero because of the dependence on 
exponential functions. The contribution from the circular cylinder 
wall is small and would be totally negligible with more data points 
leading to an extension of the galactic model to very large $r$. The 
key element to consider is the lower surface of the cylinder. The 
\textit{outward} normal is ${\bf n}={\bf -e}_z$. 

From (\ref{Eq40r}), we have
\eqnn{Eq41r}{
	M_{upper}
	=
	\int B(r, \epsilon) {\bf e}_z\cdot(-{\bf e}_z) dS
	=
	-B(\epsilon)S 
}
where $S$ is the lower surface area and $z=+\epsilon$ on the lower face. 

Similarly, for the lower cylinder, we have 
\eqnn{Eq42r}{ 
	M_{lower} = + B(-\epsilon)S
}
where $z=-\epsilon$ on the upper face of the lower cylinder.
By symmetry, $M_{upper}$ = $M_{lower}$
And since $\rho$ is positive over the continuum, it follows that
\eqnn{Eq43r}{
	B(\epsilon)<0,\quad
	B(-\epsilon)>0,\quad
	B(-\epsilon)= - B(\epsilon)
}
Now we perform the standard pill box calculation to deduce the mass 
of the surface layer when we do not impose the physical requirement 
that the density be taken as the limit value as in (\ref{Eq30r}). The 
upper surface of the pill box (see Figure \ref{fig:cylinders}) is the same as the lower 
surface of the upper cylinder and this has unit outward normal ${\bf 
+e}_z$. Hence the mass contribution is $B(\epsilon)S$ which is 
negative. Similarly, the lower surface of the pill box gives a 
contribution $-B({\bf -\epsilon})S$ which, by (\ref{Eq4r}) is also 
negative. Thus, the surface layer mass is deduced to be the negative 
of the continuum mass apart from a small contribution at the cylinder wall.

\end{document}